\documentclass[aps,twocolumn,showpacs,nofootinbib,floatfix]{revtex4-1}
\pdfoutput=1
\usepackage{bm}
\usepackage[pdftex]{graphicx}
\usepackage{multirow}
\usepackage{amsmath}
\usepackage{amssymb}
\usepackage{hyperref}
\usepackage{color}
\usepackage{xspace}
\usepackage{verbatim}
\usepackage{mathtools}
\usepackage{enumitem}
\usepackage{booktabs}
\usepackage[dvipsnames]{xcolor}
\usepackage[caption=false]{subfig}

\newcommand{\kms}{\,\,\mathrm{km}\,\mathrm{s}^{-1}}

\newcommand{\No}{N_\mathrm{o}}
\newcommand{\Ne}{N_\mathrm{e}}

\newcommand{\lpM}{\lambda_p^M}
\newcommand{\lnM}{\lambda_n^M}

\begin{document}
\title{Prospects for determining the particle/antiparticle nature of WIMP dark matter with direct detection experiments}

\author{Bradley J. Kavanagh}
\email{bradley.kavanagh@lpthe.jussieu.fr}
\affiliation{Laboratoire de Physique Th\'eorique et Hautes Energies, CNRS, UMR 7589, 4 Place Jussieu, F-75252, Paris, France}
\author{Farinaldo S.\ Queiroz}
\email{farinaldo.queiroz@mpi-hd.mpg.de}
\affiliation{Max-Planck-Institut f\"ur Kernphysik, Postfach 103980, 69029 Heidelberg, Germany}
\author{Werner Rodejohann}
\email{werner.rodejohann@mpi-hd.mpg.de}
\affiliation{Max-Planck-Institut f\"ur Kernphysik, Postfach 103980, 69029 Heidelberg, Germany}
\author{Carlos E.\ Yaguna}
\email{carlos.yaguna@uptc.edu.co}
\affiliation{Escuela de F\'{i}sica, Universidad Pedag\'ogica y Tecnol\'ogica de Colombia\\
Avenida Central del Norte, Tunja, Colombia}

\begin{abstract}

\noindent It was recently pointed out that  direct detection signals from at least three different targets may be used to determine whether the Dark Matter (DM) particle is different from its antiparticle. In this work, we examine in detail the feasibility of this test under different conditions, motivated by proposals for future detectors. Specifically, we perform likelihood fits to mock data under the hypotheses that the DM particle is  identical to or different from its antiparticle, and determine the significance with which the former can be rejected in favor of the latter. In our analysis, we consider 3 different values of the DM mass ($50$ GeV, $300$ GeV, $1$ TeV) and 4 different experimental ensembles, each consisting of at least 3 different targets -- Xe and Ar plus one  among the following: Si, Ge, $\mathrm{CaWO_4}$, or  Ge/$\mathrm{CaWO_4}$.  For each of these experimental ensembles and each DM mass, the expected discrimination significance is calculated as a function of the DM-nucleon couplings. In the best case scenario, the discrimination significance can exceed $\mathcal{O}(3\sigma)$ for three of the four ensembles considered, reaching $\mathcal{O}(5\sigma)$ at special values of the DM-nucleon couplings. For the ensemble including $\mathrm{Si}$, $\mathcal{O}(5\sigma)$ significance can be achieved for a range of DM masses and over a much wider range of DM-nucleon couplings, highlighting the need for a variety of experimental targets in order to determine the DM properties. These results show that  future direct detection signals could be used to exclude, at a statistically significant level, a Majorana or a real DM particle, giving a critical clue about the identity of the Dark Matter. 

%In the best case scenario, the discrimination significance can reach $\mathcal{O}(3\sigma)$ for three of the four ensembles considered, and  $\mathcal{O}(5\sigma)$ for the ensemble including  $\mathrm{Si}$, highlighting the need for a variety of experimental targets in order to determine the DM properties. These results show that  future direct detection signals could be used to exclude, at a statistically significant level, a Majorana or a real DM particle, giving a critical clue about the identity of the Dark Matter. 
\end{abstract}

\maketitle

\section{Introduction}
The identification of Dark Matter (DM) poses one of the most challenging problems in cosmology, particle and astroparticle physics \cite{Bertone:2010zza}. Robust and generally accepted solutions to the DM problem imply that a new particle provides the necessary energy density. The goal is then to determine the fundamental properties of this new particle, and to do so it must first be detected by non-gravitational means.  A promising way to detect the dark matter particle is to observe its scattering  with a target material in terrestrial detectors --  dubbed direct detection \cite{Goodman:1984dc,Drukier:1986tm}. 

In recent years, direct detection experiments have significantly improved the constraints on the DM-nucleon scattering cross section  \cite{Angloher:2015ewa,Agnese:2015nto,Agnes:2015ftt,Tan:2016zwf,Hehn:2016nll,Akerib:2016vxi,Aprile:2016swn,Amole:2017dex,Hehn:2016nll}, but no definitive signal has yet been observed. This year a new generation of direct detection experiments, with a target mass of order 1 ton, has entered into play and has already started probing new regions of the parameter space \cite{Aprile:2017iyp}, opening the possibility of observing a DM signal in the near future. Once such signals are detected, it remains the task of extracting from them, and possibly in combination with signals from other experiments, the fundamental properties of the DM particle \cite{Fox:2010bz,Pato:2010zk,Catena:2014uqa,Anderson:2015xaa,Roszkowski:2016bhs}. One of these properties, which has not received much attention, is the nature of the DM antiparticle. Is the DM its own antiparticle, as is the case for a Majorana fermion and for a real scalar or vector? Or is it a different particle, as is the case for a Dirac fermion and for a complex scalar or vector?

In a recent paper, a test that addresses precisely these questions was  proposed \cite{Queiroz:2016sxf} (hereafter QRY17). This test, which is based on direct detection data only,  requires the observation of spin-independent signals in three different targets and allows one to exclude a DM particle that is self-conjugate (i.e.~which is its own antiparticle). The crucial observation is that, for self-conjugate DM, the spin-independent scattering cross sections off nuclei depend on just two couplings, which determine the DM interaction with the proton and with the neutron. For DM that is not self-conjugate, there are instead four fundamental couplings (or more precisely three measurable parameters, as we will show), which determine the interactions of the dark matter particle and of its antiparticle with the proton and with the neutron. Thus, if signals are observed in more than two experiments with different targets, one may find that the interpretation with just two coupling parameters is inconsistent, and consequently that the DM particle can not be its own antiparticle.

In this paper we will perform a much more sophisticated analysis than that presented in QRY17, where only rough estimates were made of the experimental precision required to show the Dirac nature of DM. Here we instead simulate direct detection data from different targets based on projections for several experiments which may enter into operation in the near future. In addition, we implement a likelihood analysis, which allows us to properly combine the data from different experiments and to determine the precise statistical significance with which a DM particle which is self-conjugate can be discriminated from one which is not. We compute this \textit{discrimination significance} for different sets of possible experiments, for several values of the  DM mass and for different underlying DM-nucleon couplings. This allows us to highlight which experimental ensembles will be most effective at excluding self-conjugate DM and for which couplings this is feasible.

In the next section, we review the test proposed in QRY17, as it applies to a fermion DM particle, and introduce the basic notation to be used throughout the paper. The standard direct detection formalism is introduced in Sec.~\ref{sec:eventrate} while  the four different sets (ensembles) of experiments that are part of our analysis are presented in Sec.~\ref{sec:mock}. Section~\ref{sec:procedure} explains in detail the statistical procedure that we use to study the feasibility of the test, with the more technical details relegated to the appendices. Our main results are described in Sec.~\ref{sec:results}, in which we present the discrimination significance obtained with four different experimental ensembles, for several values of the DM mass. Finally, we discuss and summarize our key findings in Sec.~\ref{sec:discussion} and Sec.~\ref{sec:summary}.

\section{Dirac and Majorana Dark Matter}
\label{sec:DiracAndMajorana}

Here we review the test proposed in QRY17 for determining whether DM is its own antiparticle. This test works in exactly the same way for scalar, fermion or vector DM. For definiteness, then, we will consider fermion DM throughout our analysis  but it should be kept in mind that our results do not rely on such an assumption.

Our starting point is then the most general Lagrangian \cite{Belanger:2008sj} for a fermion DM particle $\chi$ coupling to nucleons $N = n,p$ in a spin-independent way, at leading order in the DM-nucleon relative velocity\footnote{Spin-independent interactions which are higher order in the DM velocity are of course possible. However, for DM speeds of $v \sim 10^{-3}$, these will typically be subdominant. We briefly discuss this further in Sec.~\ref{sec:discussion}.}: 
\begin{equation}\label{eq:L}
{\cal L}_{\rm SI}^F = \lambda_{N,e}\, \bar\psi_\chi\psi_\chi\,\bar\psi_N\psi_N
+\lambda_{N,o}\, \bar\psi_\chi\gamma_\mu\psi_\chi\,\bar\psi_N\gamma^\mu\psi_N\,.
\end{equation}
Here $\lambda_{N,e}$ and $\lambda_{N,o}$ are couplings (taken real for simplicity) of dimension $E^{-2}$ which depend on the explicit particle physics model underlying the interaction, and also take into account the 
translation of the fundamental quark-level Lagrangian to the hadronic level \cite{DelNobile:2013sia,Hill:2014yxa,Hoferichter:2016nvd}. The subscript 
$e$ ($o$) implies that the operator is even (odd) under the interchange of $\chi$ and $\bar \chi$. 
For Majorana particles, the odd terms are absent, and the cross section\footnote{in the zero-momentum transfer limit} for DM scattering off a nucleus $A$ is given by \cite{Cerdeno:2010jj}
\begin{align}
\nonumber
\sigma^M_{\rm SI}= & \frac{4\mu_{\chi A}^2}{\pi}\left[\lambda_{p,e}\,N_p+\lambda_{n,e}\,N_n\right]^2 \\
 \equiv & \frac{4\mu_{\chi A}^2}{\pi}\left[\lambda_{p}^M\,N_p + \lambda_{n}^M\, N_n\right]^2 \,.\label{eq:sigmaM}
\end{align}
Here $ \mu_{\chi A}=M_\chi M_A/(M_\chi+M_A)$ is the reduced mass of the DM-nucleus system, $N_p$ is the  number of protons and $N_n$ is the number of neutrons.
A Dirac particle would have the same cross section, with $\lambda_{N,e}$ replaced by 
$\lambda_{N,e} + \lambda_{N,o} \equiv 2 \lambda_N^D$. A Dirac \textit{antiparticle} would again have the same cross section, 
but with $\lambda_{N,e}$ replaced by $\lambda_{N,e} - \lambda_{N,o}\equiv 2 \lambda_N^{\overline D}$. 
If Dirac particles and antiparticles contribute equally to the observed DM density, as expected in the standard freeze-out scenario \cite{Kolb:1990vq},
the total cross section with nucleons is the average of the particle and antiparticle cross sections: 

\begin{align}
\nonumber \sigma_{\mathrm{SI}}^D &= \frac{4 \mu_{\chi A}^2}{\pi}\frac{1}{2} \left(  \left[\lambda_p^D N_p + \lambda_n^D N_n\right]^2 \right.\\
\nonumber & \qquad \qquad \left. +\left[\lambda_p^{\overline{D}} N_p + \lambda_n^{\overline{D}} N_n\right]^2     \right)\label{eq:dirac_xsec1} \\ 
\nonumber&= \frac{2 \mu_{\chi A}^2}{\pi} \left( (\lambda_p^{D \,2} + \lambda_p^{\overline{D}\, 2}) N_p^2 +  (\lambda_n^{D \, 2} + \lambda_n^{\overline{D}\,2}) N_n^2 \right. \\
\nonumber & \qquad \qquad \left. + 2(\lambda_p^D \lambda_n^{D} + \lambda_p^{\overline{D}} \lambda_n^{\overline{D}}) N_p N_n \right)\\
&= \frac{4 \mu_{\chi A}^2}{\pi} \left(\lambda_p^2 N_p^2 + \lambda_n^2 N_n^2 + 2 \lambda_p \lambda_n f N_p N_n\right)\,.
\end{align}
Here we have defined 
\begin{align}
\label{eq:lambdap}
\lambda_p &= \sqrt{\frac{1}{2}(\lambda_p^{D \,2} + \lambda_p^{\overline{D}\, 2})}\\
\label{eq:lambdan}
\lambda_n &= \sqrt{\frac{1}{2}(\lambda_n^{D \,2} + \lambda_n^{\overline{D}\, 2})}\\
\label{eq:f}
f &= (\lambda_p^D \lambda_n^{D} + \lambda_p^{\overline{D}} \lambda_n^{\overline{D}})/( 2\lambda_p \lambda_n)\,.
\end{align}
Thus, we can write:
\begin{align}
\label{eq:dirac_xsec2}
\sigma_{\mathrm{SI}}^D &= \frac{4 \mu_{\chi A}^2}{\pi} \left(\left[ \lambda_p N_p + \lambda_n N_n \right]^2 + 2 \lambda_p \lambda_n (f - 1) N_p N_n \right).
\end{align}
The three parameters $(\lambda_p, \lambda_n, f)$, with $f \in [-1, 1]$ are all that are needed to describe the Dirac DM-nucleus cross section given in Eq.~(\ref{eq:dirac_xsec1}). There is therefore a degeneracy between the DM-nucleon couplings $(\lambda_p^D, \lambda_p^{\overline{D}}, \lambda_n^D, \lambda_n^{\overline{D}})$, which cannot be broken by direct detection experiments. Without loss of generality, we set $\lambda_p^{\overline{D}} = 0$ throughout this work.

A comparison of Eq.~\eqref{eq:sigmaM} and Eq.~\eqref{eq:dirac_xsec2} makes manifestly clear that the scattering cross sections for Dirac and Majorana DM can have different dependences on the number of protons and neutrons in the target nucleus. This is the basis of the test proposed in QRY17; with positive signals in at least three different experimental targets, one can determine whether this dependence is consistent with a Majorana particle (as in Eq.~\eqref{eq:sigmaM}) or a Dirac particle (as in Eq.~\eqref{eq:dirac_xsec2}).

Notice that for $f=\pm 1$ the Dirac cross section in Eq.~\eqref{eq:dirac_xsec2} takes the form $\sigma_{\mathrm{SI}}^D \propto \left[\lambda_p N_p\pm \lambda_n N_n\right]^2$, which is equivalent to the Majorana cross section (Eq.~(\ref{eq:sigmaM})) through the identification $\lambda_p^M=\lambda_p$ and $\lambda_n^M=\pm\lambda_n$. The Majorana cross section can thus be recovered as a special case of the Dirac one with the implication that in this case direct detection experiments could never determine, even in principle, that the dark matter is a Majorana particle. For $f \neq \pm 1$, however, experiments could exclude this possibility, establishing the DM as a Dirac particle.

Let us briefly describe some special cases where $f$ happens to be equal to $\pm1$ and Dirac and Majorana DM cannot be distinguished.  The first case occurs when the DM fermion has only scalar or vector interactions, i.e.\ $\lambda_{p,e} = \lambda_{n,e} = 0$ or $\lambda_{p,o} = \lambda_{n,o} = 0$; both types of interactions should consequently be present to allow any discrimination. The second case occurs when the cross section of the particle or the antiparticle vanishes ($\lambda_{p,e} + \lambda_{p,o} = \lambda_{n,e} + \lambda_{n,o} = 0$ or  $\lambda_{p,e} - \lambda_{p,o} = \lambda_{n,e} - \lambda_{n,o} = 0$).  The value $f = \pm1$ is also obtained  when the ratio between the coupling to the proton and to the neutron is the same for the DM particle and the antiparticle: $\lambda_n^{D}/\lambda_p^{D} = \lambda_n^{\overline{D}}/\lambda_p^{\overline{D}}$. A final example is  a DM particle that couples only to protons or only to neutrons,  $\lambda_{n,e} = \lambda_{n,o}=0$ or  $\lambda_{p,e} = \lambda_{p,o}=0$, which leads to $\lambda_n = 0$ or $\lambda_p = 0$, respectively.  Thus, over the multi-dimensional parameter space of the DM couplings, which consists of  $(\lambda_p^D, \lambda_p^{\overline{D}}, \lambda_n^D, \lambda_n^{\overline{D}})$, there exists a number of \emph{special} regions where the test proposed in QRY17 is inconclusive from a theoretical point of view and that independent of the available experimental data. Outside those regions -- that is, over most of the parameter space -- the test is in principle feasible, but it may be limited by the targets that can realistically be used in  direct detection experiments and by the precision that can be reached in such experiments. These issues will be analyzed in detail in this work.

A difficulty already observed in QRY17 is that the three targets required to exclude a Majorana DM particle must differ in their ratios $N_p/N_n$ (number of protons/number of neutrons). However, this quantity does not vary much for stable nuclei. As a result, the discrepancy between $\sigma_{\mathrm{SI}}^M$ and $\sigma_{\mathrm{SI}}^D$ tends to be small and  can often be accounted for by the uncertainties on the measured cross sections. 
As discussed in Appendix~\ref{app:Analytic}, this is not necessarily the case in regions of the parameter space  where there is a partial cancellation between the proton and neutron contributions to the DM cross section off a given target. It is in such regions where the discrimination sensitivity will be maximized. This partial cancellation occurs, according to Eq.~\eqref{eq:dirac_xsec2}, when $f$ is close to $-1$ and when $\lambda_n/\lambda_p \approx N_p/N_n$ for one of the experimental targets. In this work, we map out this parameter space more precisely, by quantifying the statistical significance with which Dirac and Majorana DM can be discriminated as a function of $\lambda_p$, $\lambda_n$ and $f$. 

\section{Direct Detection Event Rate}
\label{sec:eventrate}

In order to put the method of the previous section into practice, we must first set out the formalism for calculating the event rate in direct detection experiments, from which the DM-nucleon couplings are to be estimated. The expected rate of nuclear recoils $R$ per unit nuclear recoil energy $E_R$ is obtained by convolving the DM flux with the DM-nucleus differential cross section $\mathrm{d}\sigma_{\chi A}/\mathrm{d}E_R$ \cite{Jungman:1995df}
\begin{align}
\frac{\mathrm{d}R}{\mathrm{d}E_R}&= \sum_{A}^\mathrm{nuclei} X_A\frac{\rho_\chi}{m_\chi m_A} \int_{v_\mathrm{min}}^{v_\mathrm{esc}} v f(v) \frac{\mathrm{d}\sigma_{\chi A}}{\mathrm{d}E_R}\,\mathrm{d}v\,,
\label{eqrate}
\end{align}
Here, we have allowed for the possibility that the detector is composed of several different nuclei with mass fractions $X_A$. 

We fix the local DM density to the canonical value of $\rho_\chi = 0.3 \,\,\mathrm{GeV}\, \mathrm{cm}^{-3}$, though we note that observational estimates are typically in the range 0.2--0.8 $\mathrm{GeV}\, \mathrm{cm}^{-3}$ (for a review, see Ref.~\cite{Read:2014qva}). We assume that the local DM population is well described by the Standard Halo Model (SHM), leading to an isotropic Maxwell-Boltzmann speed distribution $f(v)$ (see for instance Eq.~(18) of Ref.~\cite{Peter:2013aha}). We assume a speed dispersion $\sigma_v = 156 \kms$ and take the relative speed of the Earth with respect to the halo as $v_\mathrm{Earth} = 232 \kms$ \cite{Feast:1997sb,Bovy:2012ba,McCabe:2013kea}, which we assume constant. The speed distribution (in the Galactic frame) is truncated at the local escape speed of the Milky Way $v_\mathrm{esc} \approx 544 \kms$ \cite{Smith:2006ym,Piffl:2013mla}. We integrate over all speeds $v > v_\mathrm{min}$, the minimum DM speed required to produce a nuclear recoil with energy $E_R$:
\begin{align}
v_\mathrm{min}(E_R) &= \sqrt{\frac{m_AE_R}{2\mu_{\chi A}^2}}\,.
\end{align}

For spin-independent (SI) interactions, the differential DM-nucleus cross section can be written \cite{Cerdeno:2010jj}:
\begin{align}
\frac{\mathrm{d}\sigma_{\chi A}}{\mathrm{d}E_R} &= \frac{m_A }{2\mu_{\chi A}^2v^2} \sigma_{A} F^2(E_R)\,,
\label{eqdCS}
\end{align}
where $F^2(E_R)$ is the standard Helm form factor \cite{Helm:1956zz, Lewin:1995rx} and $\sigma_{A}$ is the DM-nucleus cross section at zero momentum transfer. The exact form of $\sigma_{A}$ is given in Eq.~\eqref{eq:sigmaM} for Majorana DM and in Eq.~\eqref{eq:dirac_xsec1} for Dirac DM (taking into account the averaging over particles and antiparticles). 

The total number of expected signal events in a given detector is then obtained by integrating over all recoil energies in the analysis window of the experiment,
\begin{align}
N_e = MT \int_{E_\mathrm{min}}^{E_\mathrm{max}} \epsilon(E_R) \frac{\mathrm{d}R}{\mathrm{d}E_R} \,\mathrm{d}E_R\,.
\label{eqevents}
\end{align}
where $MT$ is the total exposure (mass $\times$ exposure time) and $\epsilon(E_R)$ is the detector efficiency at energy $E_R$. Details of the detector properties assumed in this work are given in the next section.

\section{Mock Experiments}
\label{sec:mock}

\begin{table}[t!]\centering
\begin{tabular}{@{}lllll@{}}
\toprule
\toprule
 Target		&  $E_\mathrm{min}$ [keV]	& $E_\mathrm{max}$ [keV] & Exposure [ton yr]  &  Refs.   \\
 \hline
 Xe 	& 5 & 40 & 20 & \cite{Akerib:2015cja,Akerib:2015rjg, Akerib:2016vxi,XENONnT}\\
Ar & 30 & 200 & 150 &  \cite{Amaudruz:2014nsa,Agnes:2015ftt,DEAP-50T}\\
Ge & 5 & 100 & 3  & \cite{Angloher201441}\\
CaWO$_4$ & 10 & 100 & 3  & \cite{Angloher201441}\\
Si & 7 & 100 & 3& \cite{Angloher201441,Agnese:2013rvf}\\
\bottomrule
\bottomrule
\end{tabular}
\caption{\textbf{Mock experiments considered.} In all cases, we assume a nominal efficiency of 70\%, which should be considered as the product of the signal detection efficiency and the duty cycle of the detectors.}
\label{tab:experiments}
\end{table}

	In order to study the power of future experiments to discriminate between Dirac and Majorana Dark Matter we consider the five  mock experiments detailed in Table~\ref{tab:experiments}. They are largely  based on proposed experiments which can be expected to be taking data and releasing results during the period 2020-2025. Each mock experiment is described by the range of recoil energies used for the analysis $E_R \in [E_\mathrm{min}, E_\mathrm{max}]$ and the total exposure in ton-years. We assume a constant (energy-independent) signal efficiency of 70\% for all experiments and zero backgrounds. In some cases, this assumption is reasonable; the Argon-based DarkSide-50 detector \cite{Agnes:2015ftt}, for example, has demonstrated background-free capabilities. In other cases, this assumption will be overly optimistic, but allows us to explore a `best-case' scenario, without reference to the final background properties of a given detector.
	
	For the Xe detector, we take the exposure from the XENONnT proposals \cite{XENONnT} ($\sim$ 6 ton $\times$ 3 yr). As with all the experiments we consider, the energy threshold is hard to predict as it will depend on the final detector performance once operational. We therefore estimate realistic benchmark values for each experiment. For the Xe detector, we choose a value intermediate between the LZ conceptual design report \cite{Akerib:2015cja} and the LUX 2015 analysis \cite{Akerib:2016vxi}. The DARWIN proposal \cite{Aalbers:2016jon} for an `ultimate DM detector' will provide an even larger exposure than XENONnT. However, as we will show, the discrimination power is driven mostly by the variety of targets in use, so we will not consider this larger Xenon exposure here.
	
	For the Ar detector, we take the exposure from DEAP-50T \cite{DEAP-50T} (50 ton $\times$ 3 yr) and the approximate threshold from DarkSide-50 \cite{Agnes:2015ftt}. 
	
    	For Ge and CaWO$_4$, we take the EURECA phase 2 \cite{Angloher201441} proposals. In Tab.~\ref{tab:experiments}, we assume that the full EURECA target mass is accounted for by one or the other target. In reality, the plan is for a 50:50 mass split, which is accounted for in one of the experimental ensembles listed below (ensemble D). 
    
    \begin{table}[t!]
\begin{tabular}{@{}lrrc@{}}
\toprule
\toprule
Nucleus & $A$ & $Z$ & $N_p/N_n$\\
\hline
 Silicon (Si) & 28 & 14 & 1.0\\
 Oxygen (O) & 16 & 8 & 1.0\\
 Calcium (Ca) & 40 & 20& 1.0\\ 
 Argon (Ar) & 40& 18 & 0.82\\
 Germanium (Ge) & 73 & 32 & 0.78\\
 Xenon (Xe) & 131& 54 & 0.70 \\
 Tungsten (W) & 184 & 74& 0.67\\ 
 \bottomrule
\end{tabular}
\caption{\textbf{Composition of target nuclei.} Summary of the (approximate) atomic mass $A$, atomic number $Z$ and proton-to-neutron ratio $N_p/N_n$ for the target nuclei considered in this work.}
\label{tab:nuclei}
\end{table}
    
	For the case of a Si experiment, we take the energy thresholds from the CDMS-II Silicon detectors \cite{Agnese:2013rvf}. We consider an exposure similar to that of EURECA phase 2 (1 ton $\times$ 3 yr), despite the fact that the EURECA project \textit{does not} include plans for a  Si detector.  The reason we include in our analysis an experiment which is not currently planned is that, as observed in QRY17, Si seems to be ideal for our purposes given that its ratio $N_p/N_n=1$ is significantly different from the other targets, which are summarized in Tab.~\ref{tab:nuclei}. Note that for numerical reasons (see Appendix~\ref{app:scans}), we assume each target nucleus is composed of a single isotope. As we show later, experiments with a relatively wide range of $N_p/N_n$ are required to allow good discrimination, so we do not expect small variations in $A$ between different isotopes to have a large impact on our results. Indeed, we have checked explicitly that this approximation leads only to an $\mathcal{O}(10\%)$ shift in the discrimination significance.
    
    Let us also note here that there are also proposals for direct detection experiments based on nuclear emulsions (see e.g.~Ref.\ \cite{Aleksandrov:2016fyr}). Among the target elements in that case are C, O and H, which (to good precision) have $N_p/N_n = 1$ in their natural abundance, and would thus also be good candidates to contribute to the test of DM self-conjugacy studied here.

To discriminate Dirac from Majorana DM, data from at least three different targets is required. In our analysis, we will examine 4 different ensembles of mock experiments:
\begin{description}[itemindent=1pt]
\item[Ensemble A] Xe + Ar + Si,
\item[Ensemble B] Xe + Ar + Ge,
\item[Ensemble C] Xe + Ar + CaWO$_4$,
\item[Ensemble D] Xe + Ar + 50\% Ge + 50\% CaWO$_4$.
\end{description}
All ensembles include Xe and Ar because currently they are the most promising large scale targets for the detection of a DM signal. Let us emphasize that ensemble D corresponds to the combination of the XENONnT, DEAP-50T and EURECA phase 2 experiments\footnote{The `50\%' indicates that for Ge and CaWO$_4$ we take 50\% of the nominal exposure given in Tab.~\ref{tab:experiments}.} and is therefore closest  to the current plans for future detectors. 
\section{Statistical procedure}
\label{sec:procedure}

For a given experimental ensemble, we want to evaluate the \textit{median} (Dirac vs.~Majorana) discrimination significance which can be achieved for a range of underlying DM parameters, specified by ($m_\chi$, $\lambda_p^D$, $\lambda_p^{\overline{D}}$, $ \lambda_n^D$, $\lambda_n^{\overline{D}}$). In fact, as discussed in Sec.~\ref{sec:DiracAndMajorana}, we are free to set $\lambda_p^{\overline{D}} = 0$ without loss of generality. We also fix the overall normalization of the couplings to lie just below the final LUX bounds \cite{Akerib:2016vxi}. In practice, we choose the couplings to give a total DM-Xenon cross section which is equivalent to a standard isospin conserving DM-proton cross section of $10^{-46} \,\,\mathrm{cm}^2$ at a DM mass of 50 GeV. This fixes the number of expected DM signal events in our mock Xenon experiment ($\sim 315$ events, which we keep the same for all DM masses) and thus ensures that the LUX bounds are always respected\footnote{During the preparation of this work, the first results of the XENON1T experiment were released \cite{Aprile:2017iyp}. The benchmark cross sections used in this work are still compatible with the XENON1T bounds at approximately the 95\% confidence level.}. We also verify that bounds from Ar- and Ge-based experiments are not exceeded \cite{Agnes:2015ftt,Agnese:2015ywx}.

At a given mass, each input parameter point can then be specified by just two parameters: $\lambda_n/\lambda_p$ and $f$. At each parameter point, we generate a set of mock data for the experimental ensemble under consideration. We then calculate the maximum likelihood of obtaining the data under two hypotheses:
\begin{description}[itemindent=1pt]
\item[$\mathbf{H}_M$] Majorana-like DM, with free parameters: \\ $ \Theta = (m_\chi, \lambda_p, \lambda_n, f=\pm1)$,
\item[$\mathbf{H}_D$] Dirac-like DM, with free parameters:  \\ $\Theta = (m_\chi, \lambda_p, \lambda_n, f \in [-1, 1])$.
\end{description}

We use a background-free extended likelihood which for experiment $k$ is given by:
\begin{align}
\begin{split}
\mathcal{L}_k(\Theta) =\frac{\mathrm{e}^{-\Ne (\Theta)}}{\No !}\Ne(\Theta)^{\No}  \prod_{i=1}^{\No} P(E_R^{(i)}|\Theta)\,,
\end{split}
\end{align}
where $\No$ is the number of observed events in experiment $k$, with recoil energies $\{E_R^{(1)},...,E_R^{(\No)}\}$. Given the parameters $\Theta$, $\Ne$ is the total number of expected events in experiment $k$ and $P(E_R, \Theta)$ is the probability of measuring an event of energy $E_R$. The full likelihood is then the product over all of the experiments considered:
\begin{align}
\begin{split}
\mathcal{L}(\Theta) = \prod_{k}^{N_\mathrm{expt}} \mathcal{L}_k(\Theta) \,.
\end{split}
\end{align}
The maximum likelihood under each hypothesis was determined by sampling the parameters $(\lambda_p, \lambda_n, f)$ on a grid. The procedure is described in detail in Appendix~\ref{app:scans}. We note that the likelihood can be highly multimodal with pronounced degeneracies, so calculation of the maximum likelihood is non-trivial. We have made the code for calculating the maximum likelihood (and analysing the results) available online \cite{AntiparticleCode}.

Once obtained, we compare the maximum likelihood under the two hypotheses, $\hat{\mathcal{L}}_M$ and $\hat{\mathcal{L}}_D$, by constructing the test statistic:
\begin{align}
q = -2 (\log\hat{\mathcal{L}}_M - \log\hat{\mathcal{L}}_D)\,.
\end{align}
Under the hypothesis $\mathbf{H}_M$, the test statistic $q$ should be asymptotically \textit{half chi-square} distributed \cite{Cowan:2010js} with one degree of freedom\footnote{This one degree of freedom corresponds to the one extra free parameter under $\mathbf{H}_D$, namely $f$.}. This allows us to calculate a $p$-value for the observed value of $q$ and hence determine the significance with which $\mathbf{H}_M$ can be rejected in favour of $\mathbf{H}_D$.

For each parameter point, we generate 100 mock data sets and calculate the discrimination significance for each one. This accounts for the effects of Poisson noise and allows us to determine the \textit{median} discrimination significance expected in future experiments (i.e.\ the significance we would expect to achieve in at least 50\% of realisations). 

As already discussed in Sec.~\ref{sec:DiracAndMajorana}, discrimination between Dirac and Majorana particles is expected to be maximised when there is some partial cancellation in the cross section for DM scattering of one of the target nuclei. In Appendix~\ref{app:Analytic}, we estimate analytically which values of the DM couplings will allow for significant discrimination. With this in mind, we restrict ourselves to the following range of input parameter values: $\lambda_n/\lambda_p \in [0.5, 1.0]$ and $f \in [-1.00, -0.94]$. For a given mass and experimental ensemble, we calculate the median discrimination significance over a grid in these input couplings.
\section{Results}
\label{sec:results}
\begin{figure*}[t]
\centering
\includegraphics[width=\textwidth,]{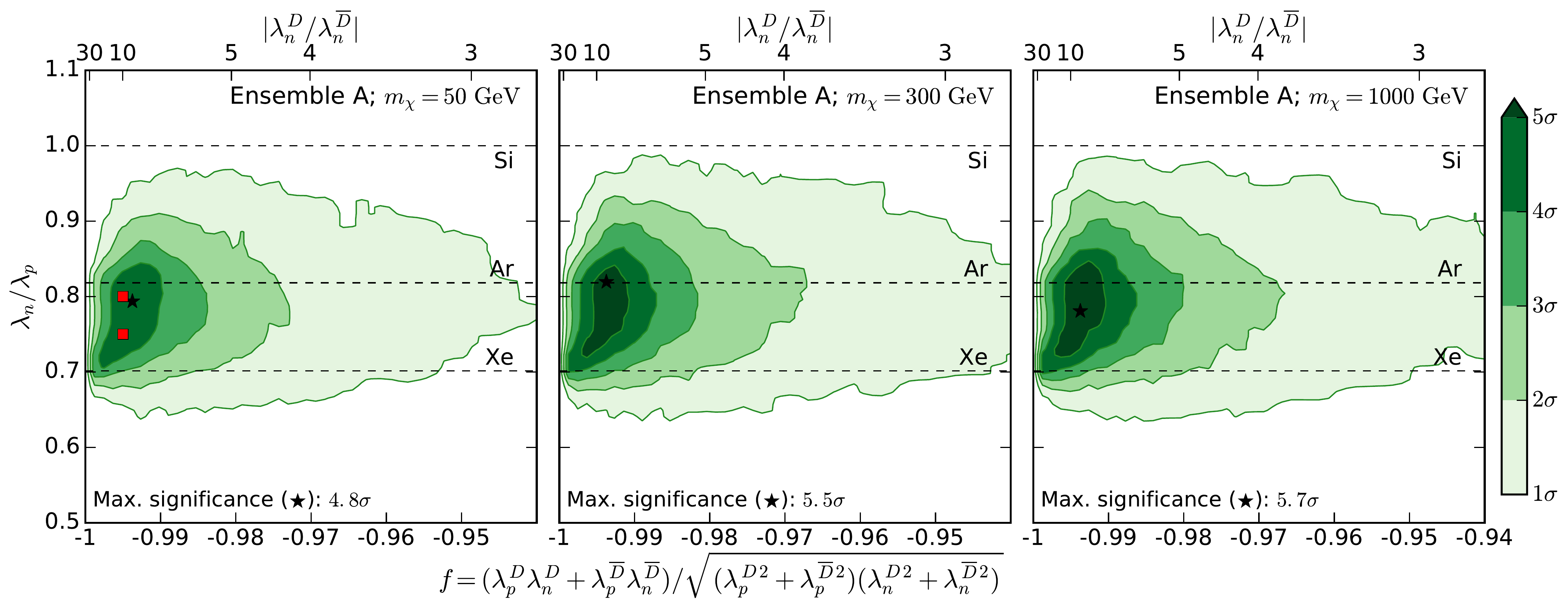}
\caption{\textbf{Median significance for discriminating Dirac from Majorana DM using ensemble A (Xe + Ar + Si).} Significance with which Dirac DM can be distinguished from Majorana DM using experimental ensemble A. Each panel shows the results for a different DM mass: 50 GeV, 300 GeV and 1000 GeV (from left to right). Dashed horizontal lines show where $\lambda_n/\lambda_p = N_p/N_n$ for the different targets in the ensemble. Along these lines (and close to $f = -1$), we expect maximal cancellation of the DM-nucleus cross section for each nucleus respectively. The parameter point with maximum discrimination significance is marked with a star. The parameters $\lambda_p$ and $\lambda_n$ are defined in Eq.\ (\ref{eq:lambdap}) and Eq.\ (\ref{eq:lambdan}). The red squares in the left panel denote the parameter points which are examined further in Fig.~\ref{fig:exposure}.}
\label{fig:ensembleA}
\end{figure*}

\begin{figure*}[tbh!]
\centering
\includegraphics[width=\textwidth,]{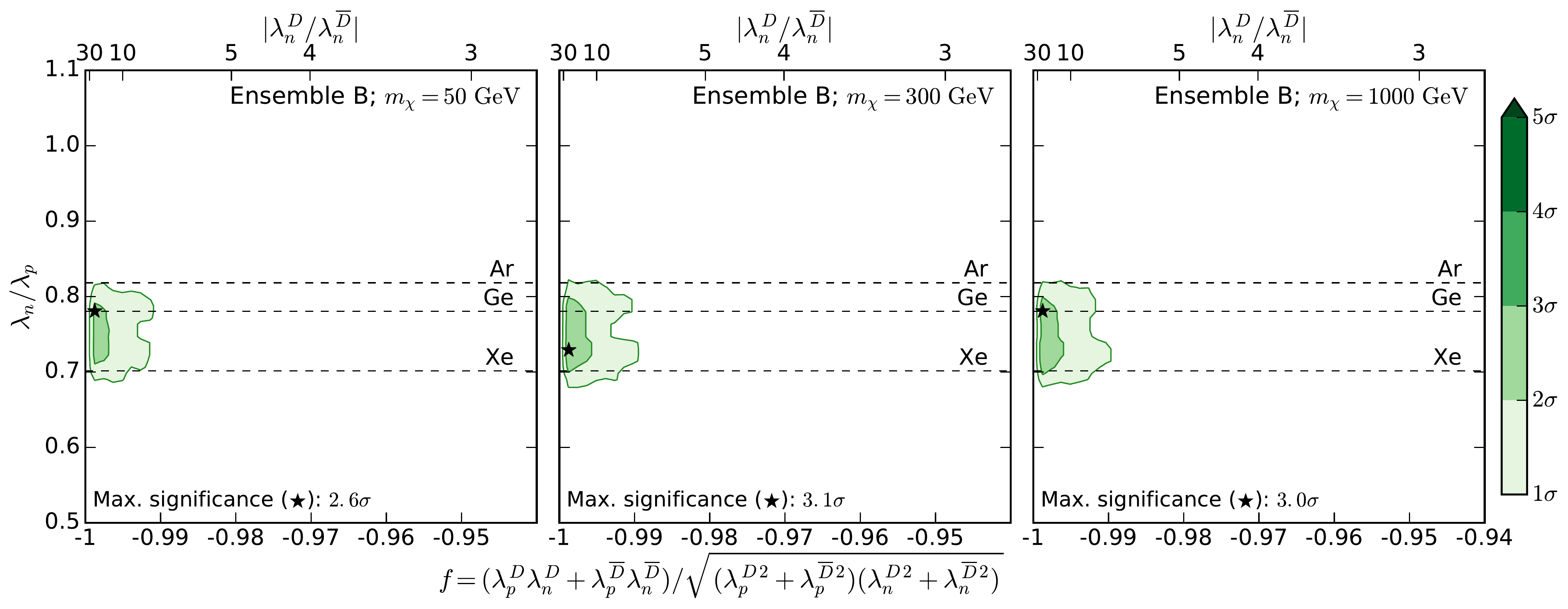}
\caption{\textbf{Median significance for discriminating Dirac from Majorana DM using ensemble B (Xe + Ar + Ge)}. As Fig.~\ref{fig:ensembleA}, but for ensemble B.}
\label{fig:ensembleB}
\end{figure*}

\begin{figure*}[tbh!]
\centering
\includegraphics[width=\textwidth,]{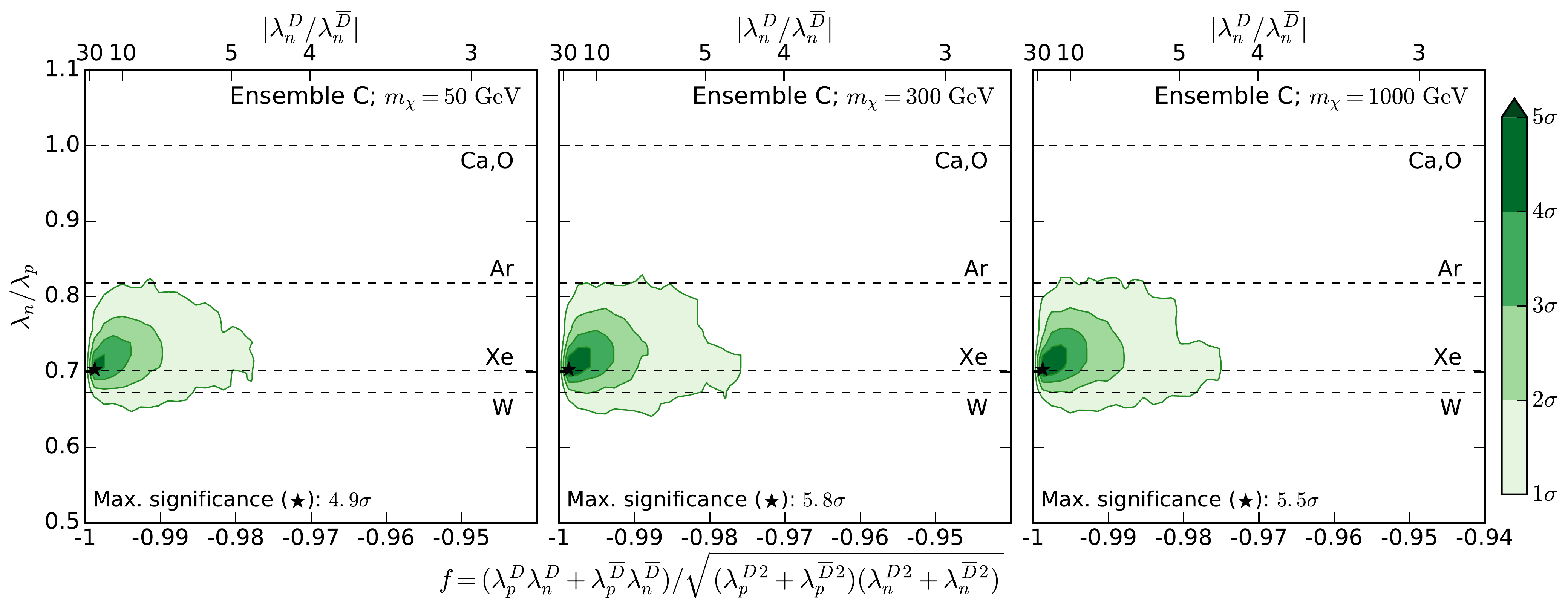}
\caption{\textbf{Median significance for discriminating Dirac from Majorana DM using ensemble C (Xe + Ar + CaWO$_4$)}. As Fig.~\ref{fig:ensembleA}, but for ensemble C.}
\label{fig:ensembleC}
\end{figure*}

\begin{figure*}[tbh!]
\centering
\includegraphics[width=\textwidth,]{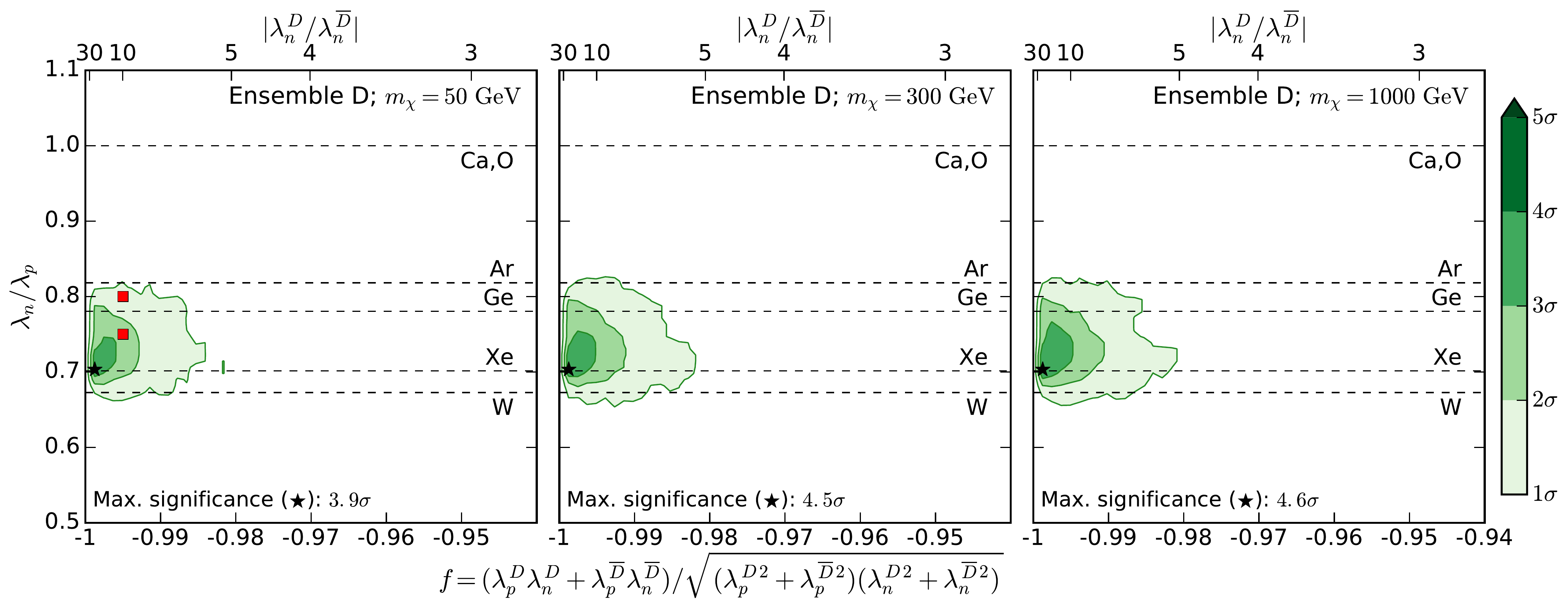}
\caption{\textbf{Median significance for discriminating Dirac from Majorana DM using ensemble D (Xe + Ar + 50\% Ge + 50\% CaWO$_4$)}. As Fig.~\ref{fig:ensembleA}, but for ensemble D. The red squares in the left panel denote the parameter points which are examined further in Fig.~\ref{fig:exposure}.}
\label{fig:ensembleD}
\end{figure*}

Let us now display our main results. Figures \ref{fig:ensembleA}-\ref{fig:ensembleD} show, in the plane $(f,\lambda_n/\lambda_p)$, the median expected discrimination significance for each of the four experimental ensembles  we consider and for dark matter masses of $50$ GeV (left panel), $300$ GeV (middle panel) and $1$ TeV (right panel). In each panel, the value of the discrimination significance is color-coded: white regions have a discrimination significance of $<1\sigma$, with darker colors corresponding to larger significance. The point with the highest significance is marked with a star. Labeled on the upper x-axis are the values of $|\lambda_n^D/\lambda_n^{\overline{D}}|$ corresponding to a given value of $f$, assuming $\lambda_p^{\overline{D}} = 0$. The dashed horizontal lines correspond to the region where $\lambda_n/\lambda_p=N_p/N_n$ for each nucleus (see Tab.~\ref{tab:nuclei}). At the point where those lines intersect the line $f=-1$ (the left axis) the expected signal is zero for that nucleus. As we will see, the regions with high discrimination  significance are always close to one of those points.

In Fig.~\ref{fig:ensembleA} we show results for ensemble A, which consists of  (Xe+Ar+Si). From the figures we can see that the $1\sigma$ regions span a limited region of the parameter space, $-1<f<0.94$ and  $0.65<\lambda_n/\lambda_p<0.95$, with a very mild  dependence  on the dark matter mass.  Only within such regions it is possible to exclude a Majorana (or self-conjugate) DM particle. These results are in agreement with the analytic estimates of Appendix~\ref{app:Analytic} (see Fig.~\ref{fig:anal-estimate}) and we find no other regions of the parameter space where a significant discrimination is possible.

For a dark matter mass of $50$ GeV (left panel), the maximum discrimination significance is $4.8\sigma$, which is reached for $f\approx -0.995$ and $\lambda_n/\lambda_p\approx -0.8$. For $m_\chi=300$ GeV and $m_\chi=1$ TeV, the maximum discrimination significance increases to $5.5\sigma$ and $5.7\sigma$ (most likely due to the increasing number of Argon events relative to Xenon at higher masses) and the point where it is reached remains close to the horizontal dashed line for Argon (where cancellation of the cross section is expected in the Argon detector).

Figure~\ref{fig:ensembleB} displays the discrimination significance for  ensemble B, which consists of (Xe+Ar+Ge). In this case, the $1\sigma$ regions are significantly smaller, hardly extending to  $f>-0.99$. The maximum discrimination significance is found to be $3.1\sigma$, achieved for a dark matter mass of $300$ GeV. At $1$ TeV the result is similar ($3.0\sigma$) whereas it is smaller for $50$ GeV ($2.6\sigma$). For all three masses, the maximum discrimination significance is reached for $f$ very close to $-1$ and for $\lambda_n/\lambda_p$ between $0.7$ and $0.8$. The lower discrimination significance for ensemble B (compared to ensemble A) is as expected. From Tab.~\ref{tab:nuclei}, we can see that the proton-to-neutron ratios in Germanium and Argon are relatively similar, making discrimination more difficult.

In Fig.~\ref{fig:ensembleC} the results for ensemble C ($\mathrm{Xe+Ar+CaWO_4}$) are displayed. In this case, the $1\sigma$ regions are a bit wider, extending up to $f\approx-0.98$. This improvement compared to ensemble B is to be expected, owing to the wider range of nuclei in the $\mathrm{CaWO}_4$ target. The maximum discrimination significance is $5.8\sigma$ and it is reached for a dark matter mass of $300$ GeV. For $1$ TeV, the maximum discrimination significance is similar ($5.5\sigma$), whereas it is a little smaller for $50$ GeV ($4.9\sigma$). Notice from the figure that for all three masses the maximum discrimination significance is reached very close to the xenon-phobic point: $\lambda_n/\lambda_p=0.7$ and $\lambda=-1$. Because we fix the normalisation of the couplings to give a fixed number of events ($\sim 315$) in our Xenon mock detector, the xenon-phobic point corresponds to large couplings and large numbers of events in the other detectors of the ensemble. Close to this point, the DM-nucleon couplings can therefore be constrained with greater precision, allowing some discrimination between Dirac and Majorana DM. As we move away from this point, however, we see that typical discrimination significances are slightly lower, in the range 3--4$\sigma$.

\begin{figure*}[tbh!]
\centering
\includegraphics[width=0.49\textwidth]{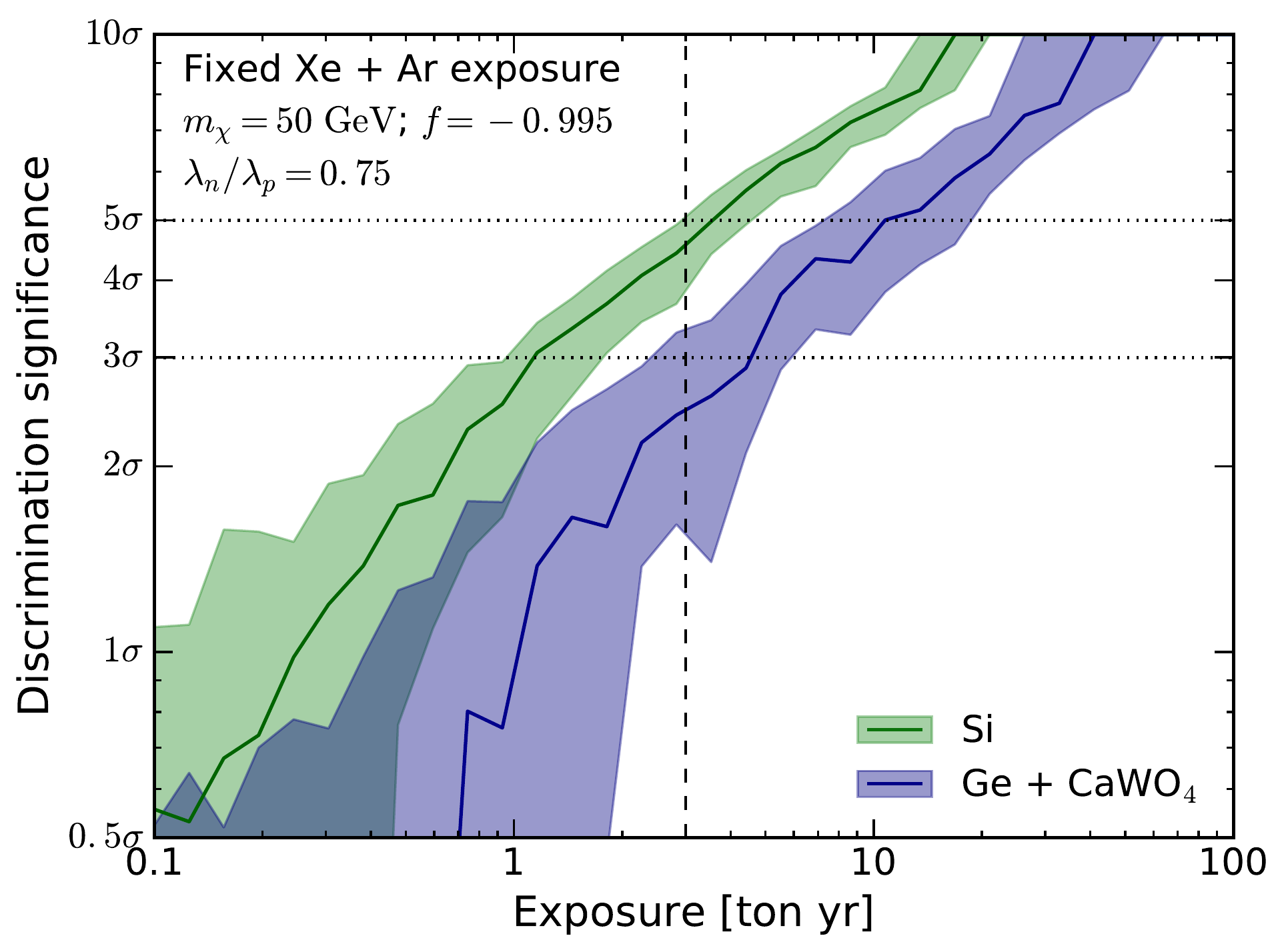}
\includegraphics[width=0.49\textwidth]{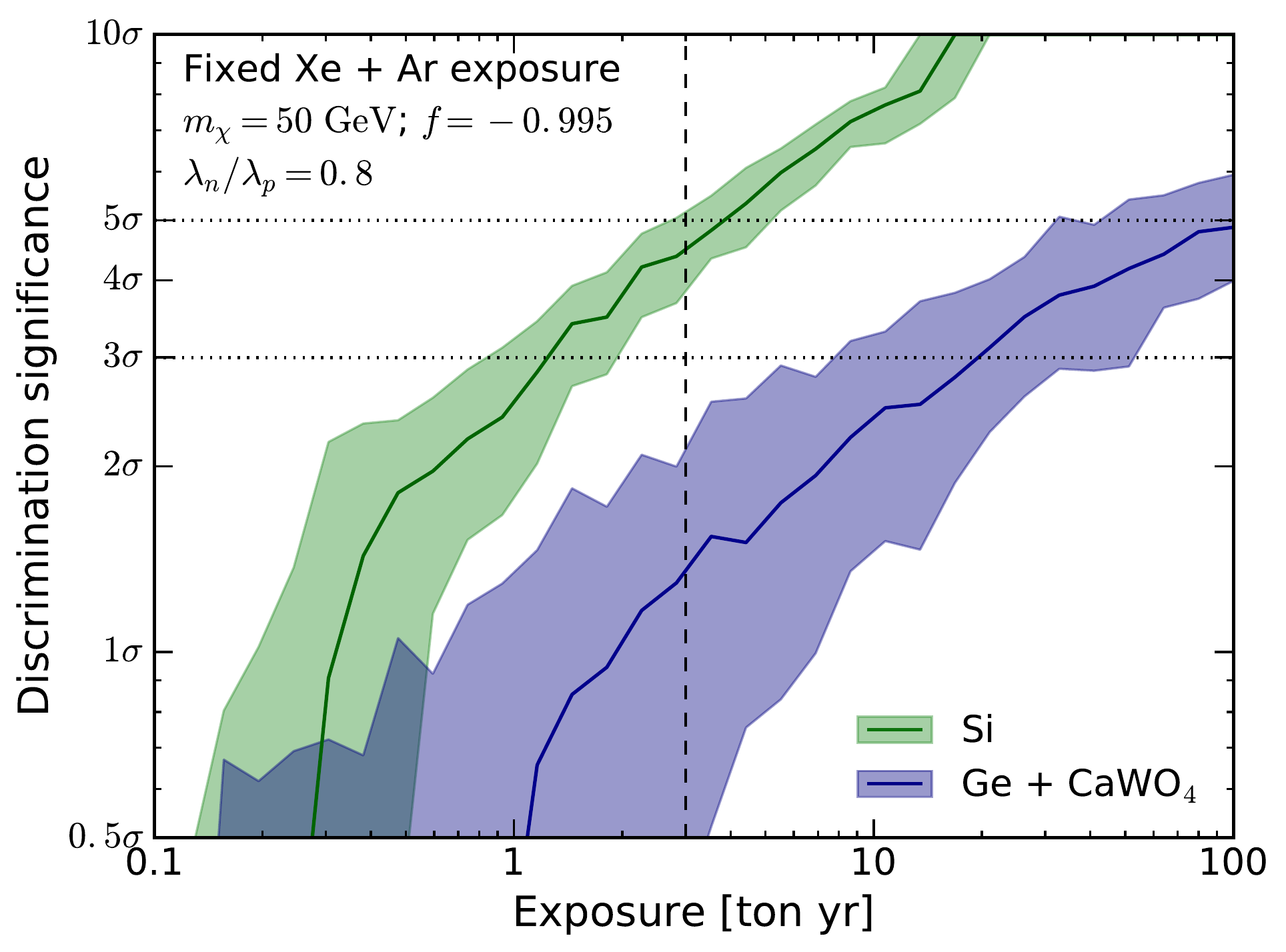}
\caption{\textbf{Median significance for discriminating Dirac from Majorana DM as a function of exposure.} Median and 68\% band for the expected discrimination significance as a function of the exposure of either an $\mathrm{Si}$ (green) or combined $\mathrm{Ge} + \mathrm{CaWO}_4$ (blue) experiment. The exposures of the Xenon and Argon detectors are fixed and given in Tab.~\ref{tab:experiments}. For the $\mathrm{Ge} + \mathrm{CaWO}_4$ experiment, the total exposure is divided equally between $\mathrm{Ge}$ and $\mathrm{CaWO}_4$. Results are for a DM mass of 50 GeV and couplings of $f=-0.995$ and $\lambda_n/\lambda_p = 0.75 \,(0.80)$ in the left (right) panel. These two parameter points are marked as red squares in the left panel of Fig.~\ref{fig:ensembleD}. The vertical dashed line corresponds to an exposure of 3 ton-years, which is the benchmark from Tab.~\ref{tab:experiments}.}
\label{fig:exposure}
\end{figure*}

Figure~\ref{fig:ensembleD} shows our results for the last ensemble in our analysis (D), which consists of ($\mathrm{Xe+Ar+Ge/CaWO_4)}$) and is perhaps the closest to current plans for future experiments.  The regions where $1\sigma$ discrimination is possible are slightly smaller than in the previous ensemble. Part of the $\mathrm{CaWO}_4$ target mass has now been traded for Ge which, as discussed, has a similar proton-to-neutron ratio as Argon and therefore makes discrimination harder. The maximum discrimination significance reaches $4.6\sigma$ for a dark matter mass of $1$ TeV, and decreases to $4.5\sigma$ and $3.9\sigma$ for $300$ GeV and $50$ GeV respectively. As in the case of ensemble C, however, we note that such high significance is only achieve very close to the xenon-phobic point. The red squares in the left panel of this  figure correspond to the parameter space points which we will examine further in Fig.~\ref{fig:exposure}.

A summary of our results for the maximum discrimination significance is presented in Tab.~\ref{tab:maximum}. In it, one can read, for each of the four ensembles we considered, the value of the maximum discrimination significance at a given DM mass. For completeness, we have included in this table also a DM mass of $25$ GeV, which was not shown in the previous figures. Note that in each case we have `maximised' over the values of $\lambda_n/\lambda_p$ and $f$. As can be seen in that table, the discrimination significance tends to be higher for heavier dark matter particles. Interestingly, we find that for ensemble B the maximum significance is of order $3\sigma$ if the dark matter mass is greater than or equal to $300$ GeV. For ensembles C and D, with the addition of CaWO$_4$, the results are more encouraging with maximum significances greater than $4\sigma$ for DM masses above 50 GeV. However, in these cases the significance drops very rapidly away from the maximum, as shown in Figs.~\ref{fig:ensembleC}~and~\ref{fig:ensembleD}. Instead, for ensemble A, which includes an Si target, a significance greater than $4\sigma$ can be achieved for all DM masses studied and indeed over a greater range of the parameter space, as shown in Fig.~\ref{fig:ensembleA}. As had been anticipated in QRY17, where a much simpler analysis was used, the observation of signals  in Xe+Ar+Si offers the best prospects for the exclusion of a Majorana (or real) dark matter particle. 

%, C and D, the maximum significance is of order $3\sigma$ if the dark matter mass is greater than or equal to $300$ GeV. For ensemble A, the results are more encouraging, as a maximum significance of order $5\sigma$ can be reached for dark matter masses above $300$ GeV. At lower masses, the maximum significance for this ensemble is still high, reaching almost $3\sigma$ at $25$ GeV and $4\sigma$ at $50$ Gev. As had been anticipated in QRY17, where a much simpler analysis was used, the observation of signals  in Xe+Ar+Si offers the best prospects for the exclusion of a Majorana (or real) dark matter particle. 

\setlength{\tabcolsep}{0.5em}
\begin{table}[t!]\centering
\begin{tabular}{@{}lllll@{}}
\toprule
\toprule
DM Mass [GeV] & 25 & 50 & 300 & 1000 \\
 \hline
A (Xe+Ar+Si) & 4.4$\sigma$ & 4.8$\sigma$ & 5.3$\sigma$ & 5.7$\sigma$\\
B (Xe+Ar+Ge)  & 2.5$\sigma$ & 2.6$\sigma$ & 3.1$\sigma$ & 3.0$\sigma$\\
C (Xe+Ar+CaWO$_4$) & 3.3$\sigma$ & 4.9$\sigma$ & 5.8$\sigma$ & 5.5$\sigma$\\
D (Xe+Ar+Ge/CaWO$_4$) & 3.1$\sigma$ & 3.9$\sigma$ & 4.5$\sigma$ & 4.6$\sigma$\\
\bottomrule
\bottomrule
\end{tabular}
\caption{\textbf{Maximum significance for discriminating Dirac and Majorana DM.} Maximum value of the median discrimination significance achievable for a range of experimental ensembles and DM masses. These values correspond to the starred points in Figs.~\ref{fig:ensembleA}-\ref{fig:ensembleD}. Note that for ensembles C and D, such high significances are only achieved in a small range of the parameter space.}
\label{tab:maximum}
\end{table}

So far, we have considered the exposure of the different experiments to be fixed according to Table \ref{tab:experiments}. It is also interesting, however, to analyze how the discrimination significance changes with the exposure at a specific parameter space point. In Fig.~\ref{fig:exposure}, we show results for ensemble A (Xe+Ar+Si, Green) as a function of the $\mathrm{Si}$ exposure and for ensemble D (Xe + Ar + 50\% Ge + 50\% CaWO$_4$, Blue) as a function of the combined $\mathrm{Ge} + \mathrm{CaWO}_4$ exposure. In both cases, we fix the Xe and Ar exposures to those given in Tab.~\ref{tab:experiments} but leave the other exposure free. We select parameter points where both ensembles are expected to achieve some discrimination: $\lambda_n/\lambda_p = 0.75, 0.80$; $f=-0.995$, and $m_\chi=50$ GeV (the red squares in Fig.~\ref{fig:ensembleD}). The vertical dashed lines correspond to an exposure of 
3 ton-years, which is the benchmark from Tab.~\ref{tab:experiments}.

As expected, the discrimination significance increases with the exposure. Even at low exposures, discrimination is much easier with the $\mathrm{Si}$ experiment. For the coupling ratio $\lambda_n/\lambda_p = 0.75$ (left panel), the gap between the performance of the two ensembles remains roughly constant. Instead, for the coupling ratio $\lambda_n/\lambda_p = 0.8$ (right panel), the gap widens, with the discrimination significance using ensemble D growing more slowly with exposure. This behaviour can be understood in the language of QRY17 \cite{Queiroz:2016sxf}, in which each experiment can be seen as providing a measurement of the Majorana DM-nucleus cross section (Eq.~\eqref{eq:sigmaM}). Each such measurement (with associated uncertainties) can then be translated into a region of parameter space in $(\lpM, \lnM)$. The data is compatible with a Majorana-like DM particle if the regions obtained from multiple experiments overlap in $(\lpM, \lnM)$. By increasing the exposure of a given experiment, we reduce the size of the region in $(\lpM, \lnM)$ which is compatible with that experiment. At some point, the size of this region becomes much smaller than the region compatible with the remaining experiments. Further increasing the exposure will not improve the discrimination substantially, as the uncertainties are driven instead by the remaining experiments (e.g.~$\mathrm{Xe}$ and $\mathrm{Ar}$). 

In the case of ensemble D, when the ratio of couplings $\lambda_n/\lambda_p = 0.8$ is close to the proton-to-neutron ratio of both Ar and Ge (right panel). There is a partial cancellation of the cross section in both Ar and Ge, meaning that the consistent regions for both experiments in $(\lpM, \lnM)$ will be roughly degenerate. After a certain point, increasing the Ge exposure does little to break the degeneracy with Ar. The increase in discrimination significance then slows, driven only by the increasing $\mathrm{CaWO}_4$ exposure. However, we emphasize that this effect does not set in until very large exposures are reached.

More quantitatively, from Fig.~\ref{fig:exposure} it can be seen that, for Si,  achieving a $5\sigma$ discrimination significance would require exposures of about 4 ton-years for both of the parameters points in the left and right panels. For the combined $\mathrm{Ge} + \mathrm{CaWO}_4$, a $3\sigma $ discrimination significance  is reached after about 5 and 15 ton-years, for $\lambda_n/\lambda_p = 0.75$ and $\lambda_n/\lambda_p = 0.8$ respectively.  These figures show that with the right combination of targets, the discrimination significance can continue to grow rapidly with exposure. This suggests that once signals are observed in direct detection experiments, there is a scientific case to keep them running beyond the 2 or 3 years that is currently the standard.

\section{Discussion}
\label{sec:discussion}

As we have seen, discriminating between Dirac and Majorana dark matter is only feasible when the DM couplings lead to partial cancellations  between the neutron and the proton contributions to the cross section off a nucleus --that is, for isospin-violating dark matter.  Isospin-violating dark matter generically denotes a scenario where the dark matter couples differently to protons and neutrons, but it is the possibility of  cancellations between their contributions that makes it particularly interesting \cite{Kurylov:2003ra, Giuliani:2005my, Feng:2011vu}. These cancellations have, in fact,  received a lot of attention over the past several years \cite{Feng:2011vu, Frandsen:2011cg, Feng:2013vaa, Feng:2013fyw}. Some explicit models for isospin-violating dark matter were studied in Refs.~\cite{Belanger:2013tla, Hamaguchi:2014pja, Martin-Lozano:2015vva, Drozd:2015gda}  while experimental constraints on these scenarios were reported (among others) in Refs.~\cite{ Gao:2011bq, Kumar:2011dr, Jin:2012jn, Hagiwara:2012we, Yaguna:2016bga}. Thus, the cancellations that are required for the test to be practical have been studied  before in other contexts and explicit models have been constructed where they take place.

The results derived in the previous section are, to a large extent, model-independent and  can, therefore, be directly applied to  any specific particle physics model of dark matter. In such  a model, the parameters $\lambda_p$, $\lambda_n$ and $f$ will not be fundamental but   would be written in terms of some characteristic couplings and mass scales.  To assess the prospects for excluding a Majorana dark matter particle in a given model, the first step would then be to determine the allowed regions for  $\lambda_p$, $\lambda_n$, and $f$,  and then to compare them with the favorable regions we found in figures \ref{fig:ensembleA}-\ref{fig:ensembleD}. The larger the  overlap between them, the better the prospects for exclusion. 

\begin{figure}[t!]
\centering
\includegraphics[width=0.5\textwidth]{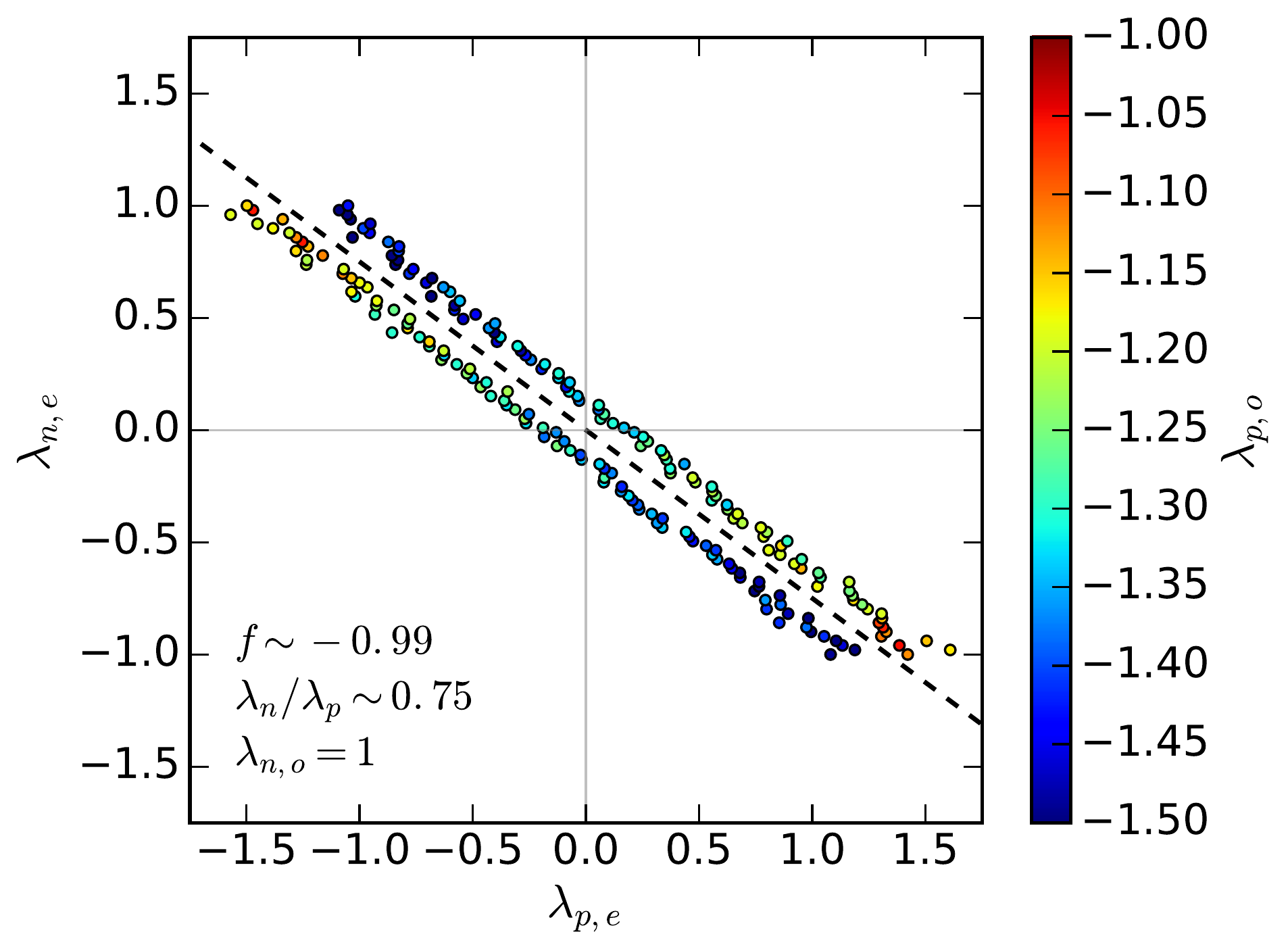}
\caption{\textbf{Relationship between Lagrangian couplings and observable parameters $f$, $\lambda_p$ and $\lambda_n$.} Random sample of points in the parameter space $\lambda_{p,e}$, $\lambda_{p,o}$, $\lambda_{n,e}$, $\lambda_{n,o}$ (see Eq.~(\ref{eq:L})) which satisfy $f \in [-0.995, -0.985]$ and $\lambda_n/\lambda_p \in [0.7, 0.8]$ (see Eqs.~\eqref{eq:lambdap}--\eqref{eq:f}). The points shown here therefore correspond to a region of parameter space where discrimination between Dirac and Majorana DM is most promising. We fix $\lambda_{n,o} = 1$ (which simply fixes the overall normalisation of the couplings) and color each point by the value of $\lambda_{p,o}$. The dashed diagonal line corresponds to $\lambda_{n,e}/\lambda_{p,e} = 0.75$. }
\label{fig:fundamental}
\end{figure}

In Fig.~\ref{fig:fundamental}, we provide an illustration of how the parameters $\lambda_p$, $\lambda_n$, and $f$ relate to the more fundamental couplings $\lambda_{N,e}$ and $\lambda_{N,o}$ which appear in the Lagrangian of Eq.~(\ref{eq:L}). For parameters in the range $f \in [-0.995, -0.985]$ and $\lambda_n/\lambda_p \in [0.7, 0.8]$ (a region where good discrimination is expected, see figures \ref{fig:ensembleA}-\ref{fig:ensembleD}), we plot the corresponding Lagrangian-level couplings (fixing $\lambda_{n,o} = 1$ in order to fix the overall normalisation). These points are aligned along two lines in the parameter space ($\lambda_{p,e}$, $\lambda_{p,o}$, $\lambda_{n,e}$) whose slope is determined by the desired ratio of couplings to protons and neutrons\footnote{There are two `allowed' lines due to an overall sign degeneracy.} For given values of $\lambda_n/\lambda_p$ and $f$, the couplings must be chosen to lie on one of these lines. However, we note that this does not require any hierarchy between the different couplings. As shown in Fig.~\ref{fig:fundamental}, it should be possible to achieve a significant discrimination with all couplings of order unity (up to some overall normalisation). This corresponds to each of the spin-independent interactions in Eq.~(\ref{eq:L}) contributing roughly equally. This also means that we do not expect the presence of subdominant (e.g.\ velocity suppressed) interactions to affect the results presented here, unless their couplings are sufficiently enhanced so as to be comparable to the standard spin-independent rate.

Let us now briefly discuss some caveats to our conclusions. On the theoretical side, our results rely on the assumption that the density of dark matter particles and antiparticles is the same. That is certainly what is expected in the standard freeze-out scenario, but it is not difficult to imagine alternatives, such as asymmetric dark matter, where it does not hold. In the more general case, one would need an additional parameter that determines the fraction of the dark matter density that is accounted for by DM antiparticles. We have also assumed that the dark matter consists of a single field with predominantly spin-independent interactions. In scenarios with multi-component dark matter (see for example Ref.~\cite{Aoki:2012ub} and references therein) or non-standard interactions (see for example Ref.~\cite{Catena:2015qad} and references therein), a more complicated analysis would be required. In such cases, we emphasize that an even greater variety of direct detection targets would likely be required to disentangle particle from antiparticle.

On the astrophysical side, we have not taken into account the uncertainties that affect the number of expected events in a given detector.  For simplicity, we considered a single fixed DM speed distribution $f(v)$, the SHM, but in reality not only are the parameters of the SHM subject to uncertainties \cite{Green:2017odb} but there are also indications that the true distribution may deviate from a smooth Maxwell-Boltzmann distribution. The evidence for such deviations from hydrodynamical simulations is reviewed in Ref.~\cite{Bozorgnia:2017brl}. A number of techniques for simultaneously fitting particle physics parameters and the local speed distribution have been developed  (see, for example, Refs.~\cite{Lee:2012pf,Kavanagh:2013wba,Feldstein:2014gza,Ibarra:2017mzt}). Incorporating such astrophysical uncertainties into the present study, we would expect the discrimination significance to be reduced, owing to a greater freedom to tune the number of events observed in each detector. The present results should therefore be taken as an optimistic case. However, we note that for relatively light DM ($m_\chi \lesssim 100\,\,\mathrm{GeV}$), using an ensemble of experiments with a range of nuclear masses should allow the speed distribution (as well as the DM mass and cross sections) to be well constrained \cite{Peter:2013aha}. In that case, we expect our results to be rather realistic.

\section{Summary}
\label{sec:summary}

In this work we investigated in detail the feasibility of distinguishing  dark matter particles that are self-conjugate (Majorana fermion and  real scalar or vector) from those that are not (Dirac fermion and complex scalar or vector) using future signals from direct detection experiments. To that end, we first simulated data from different direct detection experiments that may enter into operation in the near future.  Then, we performed fits to such data under the hypotheses that the DM is identical to or different from its antiparticle, and determine the significance with which the former can be rejected in favor of the latter. This discrimination significance was calculated, as a function of the DM couplings, for different experimental ensembles and several values of the DM mass. Our results are illustrated in 
Figs.\ \ref{fig:ensembleA}-\ref{fig:exposure} and summarized in Table \ref{tab:maximum}. The key conclusions of this study are as follows: 
\begin{itemize}

\item It is feasible to use signals from future direct detection experiments to exclude, at a statistically significant level, a Majorana or a real DM particle.  

\item Discrimination between Dirac and Majorana DM (or between real and complex DM) can be achieved only in certain regions of the parameter space. Specifically, we identified as the most promising region that one where the DM couplings lead to a partial cancellation in the DM-nucleus cross section for one of the experimental targets. That is, for $f\approx -1$ and $\lambda_n/\lambda_p\in (0.7,0.8)$ (see Eq.~\refeq{eq:lambdap}-\refeq{eq:f} for definitions). In Figs.~\ref{fig:ensembleA}-\ref{fig:ensembleD} we focused precisely on such regions.

\item According to current plans for future detectors (our ensemble D), the maximum discrimination significance that could be achieved is around 4-5$\sigma$, and depends only slightly on the dark matter mass. However, this is possible only very close to the xenon-phobic point in parameter space, dropping to roughly $3\sigma$ away from this point.

\item A Silicon target, which does not currently figure among future detectors,  could help achieve up to $5\sigma$ discrimination significance over a wider range of the parameter space, for an exposure similar to that of EURECA phase 2. We therefore propose that large-scale Silicon detectors should be considered as part of plans for future detectors such as EURECA.

\item The discrimination significance does not flatten quickly as a function of the exposure. Consequently, once direct detection signals are observed, it may be worthwhile to keep the experiments running beyond the 2 or 3 years that are currently planned. 

\end{itemize}

\acknowledgments

BJK is supported by the European Research Council ({\sc Erc}) under the EU Seventh Framework Programme (FP7/2007-2013)/{\sc Erc} Starting Grant (agreement n.\ 278234 --- `{\sc NewDark}' project). WR is supported by the DFG with grant RO 2516/6-1
in the Heisenberg Programme. 
Access to the High Performance Computing facilities provided by the Institut des Sciences du Calcul et des Donn\'ees (Sorbonne Universit\'es) is gratefully acknowledged.

\appendix

\section{Parameters scans}
\label{app:scans}

Here, we describe the procedure used to determine the maximum likelihood for each of the two hypotheses (Majorana-like or Dirac-like couplings) described in Sec.~\ref{sec:procedure}. In order to perform a large number of fits (100 scans per parameter point, over a grid of 1024 input parameter values, for several experimental ensembles), it is necessary to determine the maximum likelihood quickly and with high accuracy. We have found that Markov Chain Monte Carlo and Nested Sampling methods (with a relatively small number of samples, as required for a fast exploration of the parameter space) often fail to find the global maximum in the multi-modal likelihoods considered here. We instead sample the likelihood on a grid.

As demonstrated in Eq.~\eqref{eq:dirac_xsec2}, the spin-independent DM-nucleus cross section for Dirac DM can be described with 3 parameters: $\lambda_p$, $\lambda_n$ and $f$. The parameters $\lambda_{p,n}$ may take any positive values, but (from its definition in  Eq.~\eqref{eq:f}) we require $f \in [-1, 1]$. Note that a Majorana-like cross section is a special case of this with $f=\pm1$.

For a given DM mass, the recoil energy spectrum for scattering off a given nucleus is fixed. In this case, the log-likelihood is given by:
\begin{align}
\label{eq:loglike1}
\begin{split}
\log\mathcal{L} = - \Ne + \No \log(\Ne) + \sum_{i=1}^{\No} \log(P(E_R^{(i)}))\,,
\end{split}
\end{align}
where $P(E_R)$ does not depend on the couplings. In this case, the log-likelihood can be calculated very quickly on a dense grid over the couplings, which only affect the value of $\Ne$.\footnote{In order to be able to use Eq.~\eqref{eq:loglike1} to calculate the Xenon likelihood, we approximate the detector as containing a single isotope with $(A, Z) = (131, 54)$. This has a negligible effect on the direct detection rate, but allows us to calculate the likelihood much more quickly.} If an experiment consists of multiple targets, the likelihood is
\begin{align}
\begin{split}
\log\mathcal{L} = - \sum_{k} \Ne^k + \sum_{i=1}^{\No} \log(\sum_{k} \Ne^k P_k(E_R^{(i)}))\,,
\end{split}
\end{align}
where $\Ne^k$ is the number of expected recoils off nucleus $k$ and the sum is over all the nuclear targets in a given experiment. This is slightly more complicated but still permits a rapid calculation of the log-likelihood for a given DM mass. The full log-likelihood is then obtained by summing over experiments.

For a given mock data sample, we scan over 25 values of the DM mass, to calculate the maximum likelihood in each case (and therefore the overall maximum likelihood). For each DM mass, we calculate the log-likelihood on a grid of couplings, linearly spaced over the ranges:
\begin{align}
\label{eq:ranges}
\begin{split}
\log_{10}(\lambda_p/\mathrm{GeV}^{-2}) &\in [-10, -6]\,,\\
\log_{10}(\lambda_n/\mathrm{GeV}^{-2}) &\in [-10, -6]\,,\\
f &\in [-1, 1]\,.
\end{split}
\end{align}
In the case of Majorana-like DM, we use a grid of $(200 \times 200)$ values of $(\lambda_p, \lambda_n)$, each for $f=1$ and $f=-1$. For Dirac-like DM, we use a grid of  $(50 \times 50 \times 50)$ points in $(\lambda_p, \lambda_n, f)$.

From this initial grid scan, we obtain an estimate of the maximum likelihood points under the Dirac-like hypothesis and the Majorana-like hypothesis. For each hypothesis, we then perform 10 refinement steps, recalculating the likelihood on another grid, using the same number of grid points, but each time over a smaller range of parameter values. The range of parameters for each refinement step is centred on the maximum-likelihood parameter value from the previous step.

We repeat this procedure (scanning and refining over couplings and masses to obtain the maximum likelihoods for the Dirac-like and Majorana-like hypotheses) for each mock dataset. By generating and fitting 100 mock datasets, we estimate the median significance which can be obtained with a given ensemble of experiments.

We have verified that the method gives good convergence, i.e.\ that increasing the number of grid points does not significantly affect the results. The code used to perform the scans, along with code to analyse and plot the resulting data is publicly available and can be downloaded at \href{https://github.com/bradkav/AntiparticleDM}{https://github.com/bradkav/AntiparticleDM} \cite{AntiparticleCode}.

\section{Analytic Estimates}
\label{app:Analytic}

The likelihood-based calculations described in App.~\ref{app:scans} are computationally expensive, so it is important to obtain an initial estimate of which parts of parameter space will maximise the discrimination significance. We can then focus on these regions for the full likelihood-based calculations, rather than wasting computational time on parameter points where the significance is expected to be low.

In order to obtain this estimate, we consider having three experimental targets: $X$, $Y$ and $Z$. If DM is a Dirac particle, the DM-nucleus cross section for each target is given by Eq.~\eqref{eq:dirac_xsec1}. We write these cross sections as $\sigma^D_{X}$, $\sigma^D_{Y}$ and $\sigma^D_{Z}$. 

We imagine that experiments $X$ and $Y$ measure their respective DM-nucleus cross sections to be $\tilde{\sigma}_{X}$ and $\tilde{\sigma}_{Y}$ respectively. As described in detail in QRY17~\cite{Queiroz:2016sxf}, we can use these two measurements to estimate the DM-nucleon couplings ($\lpM$,  $\lnM$), assuming that the DM is a \textit{Majorana} particle. That is, we solve:
\begin{align}
\left[ \lpM N_p^X + \lnM N_n^X\right]^2 &= \frac{\pi \tilde{\sigma}_X}{4 \mu_{\chi X}^2}\\
\left[ \lpM N_p^Y + \lnM N_n^Y\right]^2 &= \frac{\pi \tilde{\sigma}_Y}{4 \mu_{\chi Y}^2}\,.
\end{align}
There are two possible solutions for ($\lpM$,  $\lnM$), up to an overall sign degeneracy. With these, we can calculate the DM-nucleus cross section $\sigma^M_Z$ which we would expect in a third target $Z$, assuming again a Majorana DM particle. 

We then compare $\sigma^M_Z$, the cross section expected in a $Z$ target under the assumption of a Majorana particle, and $\sigma^D_Z$, the true DM-nucleus cross section we would measure, given the Dirac nature of the particle. To do this, we evaluate the fractional difference $\Delta$ between the two cross sections,
\begin{align}
\label{eq:delta}
\Delta = \frac{(\sigma_Z^D - \sigma_Z^M)^2}{(\sigma_Z^D)^2}\,,
\end{align}
as a function of the input parameters $f$ and $\lambda_n/\lambda_p$. Small values of $\Delta$ indicate that the data should be well described by both Majorana and Dirac DM, while large values imply that the cross sections in the Majorana and Dirac scenarios should differ substantially, suggesting that significance discrimination should be possible.

\begin{figure}[t!]
\centering
\includegraphics[width=0.45\textwidth]{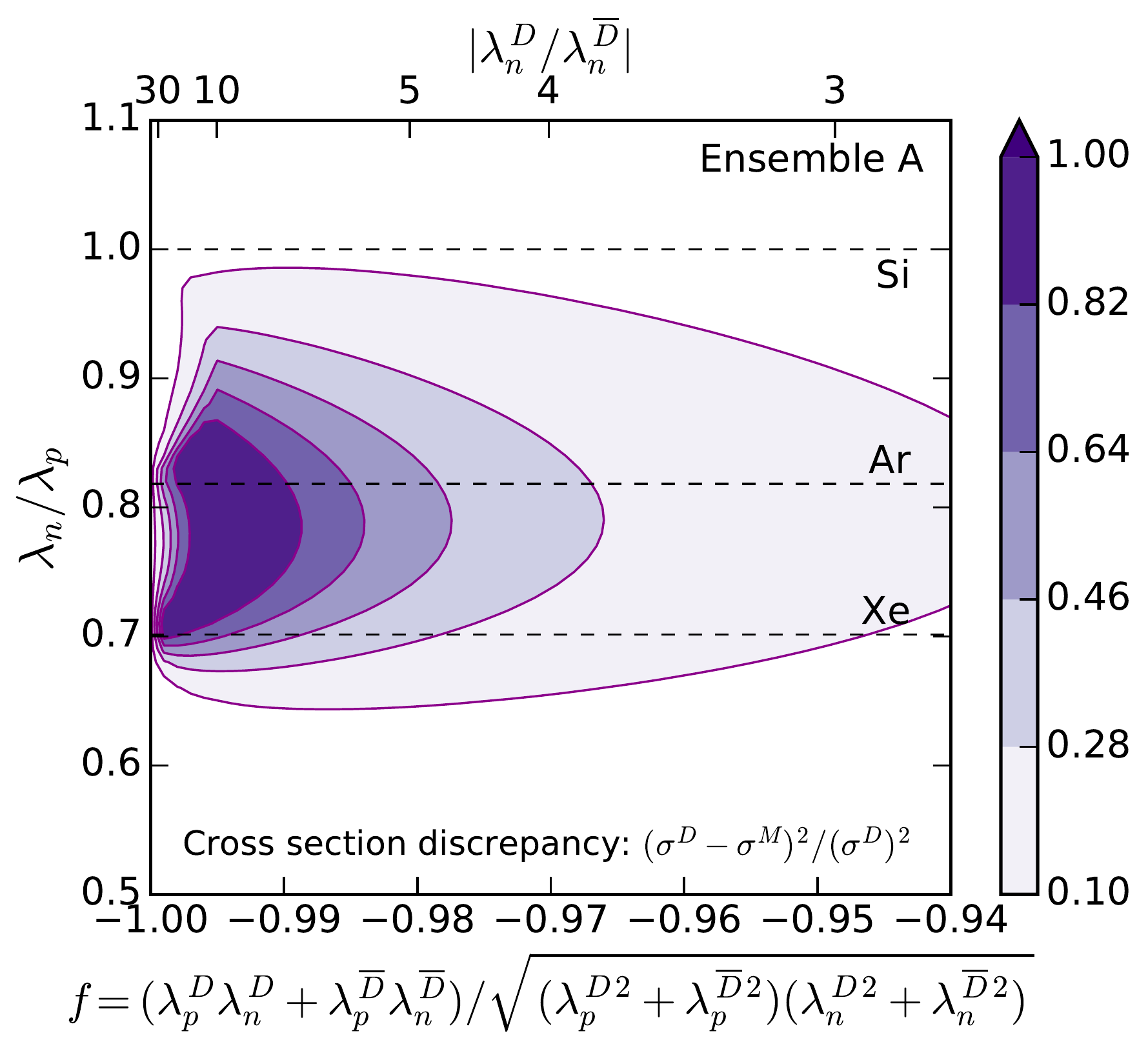}
\caption{\textbf{Analytic cross section discrepancy between Dirac and Majorana DM.} Discrepancy $\Delta$ (Eq.~\eqref{eq:delta}) between the DM-Xe cross section expected for a Majorana particle $\sigma^M$ (estimated from the `measured' DM-Si and DM-Ar cross sections) and for a Dirac particle $\sigma^D$ (using the `true' coupling values).}
\label{fig:anal-estimate}
\end{figure}

In Fig.~\ref{fig:anal-estimate}, we plot the cross section discrepancy $\Delta$ for ensemble A. Here, we have used the Ar and Si experiments to estimate the Majorana couplings and plot the value of $\Delta$ calculated for the Xe experiment. Due to the large target mass and $A^2$ coherent enhancement of Xenon-based detectors, we expect the Xe experiment to observe the largest number of events and therefore to obtain the most precise estimate of the DM-nucleus cross section. It is the discrepancy in Xenon which we therefore expect to drive the discrimination significance.

From Fig.~\ref{fig:anal-estimate}, we see that the largest discrepancies between the Majorana and Dirac cross sections are obtained when there is a partial cancellation of the DM-Xe or DM-Ar cross sections, in agreement with the discussion of Sec.~\ref{sec:DiracAndMajorana}. Instead, where there is no substantial cancellation (far from $f=-1$ or where $\lambda_n/\lambda_p$ does not match $N_p/N_n$ for any of the target nuclei) the discrepancy is smaller ($ < \mathcal{O}(10\%)$). Such a difference is likely to fall within the statistical errors of a cross section measurement and so discrimination will be difficult.

We have checked these analytic estimates for different ensembles and over much wider range of parameter values. We find in all cases that the cross section discrepancy decreases rapidly away from the parameter region depicted in Fig.~\ref{fig:anal-estimate}. We therefore focus in this work on the parameter ranges $\lambda_n/\lambda_p \in [0.5, 1.0]$ and $f \in [-1.00, -0.94]$.

\bibliography{Antiparticle}
\end{document}